# Characterization and Mitigation of Boundary Sampling Artifacts in Electron Phase Space Generation for Clinical FLASH-RT Deployment

Rafael Carballeira[1], Rongxiao Zhang[1,3], Kevin J. Willy[1], Hayley Cash[4], David J. Gladstone [1,2]

Thayer School of Engineering, Dartmouth College, Hanover, New Hampshire[1]

Dartmouth Cancer Center, Lebanon, New Hampshire[2]

School of Medicine, University of Missouri, Columbia, Missouri[3]

Medical University of South Carolina, Charleston, South Carolina[4]

## Author Contribution Statement

Rafael Carballeira performed Monte Carlo simulations, data processing, and preparation of manuscript under the advisory from Rongxiao Zhang and David J. Gladstone. Kevin J. Willy assisted in measured data collection. Hayley Cash assisted with statistical analysis and manuscript preparation. All authors have read and agreed to the published version of the manuscript.

## Conflict of Interest Statement

The authors have no conflicts of interest to disclose.

## Acknowledgments

The authors acknowledge NIH 1R21CA125760-01A1, NIH U01 CA260446-01A1, and the Norris Cotton Cancer Center Shared Resources with NCI cancer center support grant NIH 5P30 CA 23108-40. The authors graciously thank the guidance and advice provided by Dr. Jungwook Shin.

## Abstract

**Background:** Clinical implementation of Monte Carlo (MC)-based treatment planning for electron FLASH radiotherapy requires computationally efficient phase space (PHSP) generation strategies to meet the planning timelines of ultra-high-dose-rate (UHDR) delivery. Applicator-specific PHSP files, recorded at the aperture exit, reduce dose calculation time by 30–50% relative to traditional placement at the linac-exit window, enabling clinically practical planning cycles for FLASH treatment planning system (TPS) deployment. However, positioning PHSP scoring planes coincident with the applicator–air interface introduces boundary sampling artifacts that have not been previously characterized in the electron PHSP literature.

**Purpose:** To identify and characterize a boundary sampling artifact arising in Geant4-based Monte Carlo simulations when PHSP scoring planes are positioned coincident with material–air interfaces, and to demonstrate its mitigation through systematic scoring plane offset, enabling clinical deployment of applicator-specific electron beam PHSP libraries for FLASH-RT treatment planning.

**Methods:** Applicator-specific PHSP files were generated using GAMOS 6.2.0 for a commissioned 9 MeV Mobetron UHDR Monte Carlo model across twelve clinical aperture configurations (2.5–10 cm diameter) intended for deployment in the GMV Radiance treatment planning system. The scoring plane sensitivity analysis was performed on the A6 short cone (aperture exit at 397 mm) across three positions: coincident with the material-air interface (397 mm), 0.1 mm downstream (397.1 mm), and 1 mm downstream (398 mm). The corresponding 1 mm offset was subsequently applied to the A10 long cone (aperture exit at 406 mm, scoring plane at 407 mm) for global validation. Dose calculations were benchmarked against linac-exit PHSP references and experimental water phantom measurements via distance-to-agreement (DTA) analysis against the 3 mm clinical acceptance criterion established by AAPM TG-105.

**Results:** Scoring at the exact interface produced systematic proximal R50 shifts of up to 2.2 mm and DTA values of 4–6 mm for small-aperture configurations, exceeding the clinical acceptance criterion and rendering the PHSP library unsuitable for clinical deployment. A 0.1 mm offset partially corrected the angular distribution at the boundary but failed to resolve the energy-loss artifact affecting the distal falloff region. A 1 mm offset fully resolved all artifacts, achieving mean DTA values of 1.45 ± 0.63 mm (A6 series PDDs) and 2.00 ± 0.61 mm (A10 series PDDs), equivalent to or better than the linac-exit reference across all twelve configurations.

**Conclusions:** Boundary sampling artifacts in Geant4-based PHSP generation arise from the degenerate behavior of the fUseSafety step-limitation algorithm when safety = 0 at exact material boundaries, producing incomplete secondary electron equilibration and a systematic energy-loss artifact in the scored particle population that manifests as a proximal shift of the distal dose falloff. A 1 mm downstream offset fully mitigates these artifacts while introducing negligible perturbation to primary beam characteristics, enabling clinical deployment of the resulting applicator-specific PHSP library in the GMV Radiance TPS for FLASH-RT treatment planning. This offset requirement warrants investigation in other Geant4-based frameworks (including TOPAS and GATE) scoring PHSP files at applicator or collimator exit surfaces.

## 1. Introduction

Ultra-high dose rate (UHDR) electron FLASH radiotherapy (FLASH-RT) delivers dose at rates exceeding 40 Gy/s and produces biological responses that depend critically on both total dose and instantaneous dose rate[1,2]. Unlike conventional electron radiotherapy — where analytical algorithms have supported clinical intraoperative and skin treatments for decades — FLASH-RT clinical implementation requires Monte Carlo (MC)-based treatment planning because analytical dose calculation algorithms cannot reliably model the complex scattering characteristics of UHDR electron beams[3], and the dose-rate-dependent biological response requires accurate reproduction of the full beam fluence to support future dose-rate-weighted planning objectives. This requirement is particularly acute for dedicated FLASH delivery systems such as the Mobetron UHDR mobile electron linear accelerator (IntraOp Medical Corp.), where clinical deployment depends on the availability of MC-based treatment planning systems (TPS) validated for UHDR conditions.

The clinical translation of MC-based FLASH-RT planning is currently gated by computational efficiency. A central component of any MC-based TPS workflow is the phase space (PHSP) file — a stored record of particle position, direction, energy, and type at a defined scoring plane — which allows the computationally expensive accelerator head transport to be performed once and reused across many dose calculations. In conventional workflows, PHSP files are recorded at the linac exit window and replayed for each plan, but this approach still requires full particle transport through 20–40 cm of applicator geometry per calculation, typically requiring 45–60 minutes per plan for the Mobetron UHDR system corresponding to a timeline incompatible with the rapid planning cycles that clinical workflows demand. Applicator-specific PHSP files recorded immediately downstream of the field-defining aperture eliminate this redundant transport entirely, reducing calculation time by 30–50% while establishing a direct correspondence between clinical aperture selection and the computational source model. The present work is part of an active clinical deployment effort to commission the GMV Radiance TPS for Mobetron UHDR FLASH-RT, in which the applicator-specific PHSP library described here forms the foundational source model on which downstream commissioning (output factors, end-to-end testing, and independent dose verification) depends.

Implementing applicator-exit PHSP files introduces a critical and previously undescribed technical challenge: the proximity of the scoring plane to a material-air interface. In Geant4-based frameworks including GAMOS, electron transport is governed by condensed history (CH) algorithms in which the G4UrbanMscModel approximates the cumulative effect of thousands of elastic scattering events into discrete transport steps[4,5,6]. These multiple scattering (MSC) theories are mathematically valid only within infinite homogeneous media — an assumption that is necessarily violated when a step is truncated at a material boundary. At the exact interface, Geant4's internal safety parameter is defined as the isotropic distance to the nearest boundary and used by the fUseSafety step-limitation algorithm to constrain step length evaluates to zero, removing the boundary-proximity constraint from the step calculation entirely. Under this degenerate condition, low-energy secondary electrons scattered from the aperture walls are recorded before their transport has equilibrated in air, condensed-history energy-loss corrections are applied inconsistently across truncated boundary-crossing steps, and lateral displacement corrections are underestimated. The net result is a corrupted phase space distribution that produces systematic errors in downstream dose calculations.

In initial attempts to implement applicator-exit PHSP files for the Mobetron UHDR system, these artifacts manifested as premature falloff in the percent depth dose (PDD) buildup region and proximal R50 shifts of up to 2.2 mm, with DTA errors reaching 4–6 mm for small-aperture configurations. Under the 3 mm clinical acceptance criterion established by AAPM TG-105 for MC-based treatment planning[7], these errors are large enough to render the PHSP library clinically invalid and would preclude its deployment in the Radiance TPS. Characterizing and mitigating this artifact was therefore a prerequisite for clinical implementation, not an optional refinement.

This work presents a systematic investigation of these boundary sampling artifacts and their mitigation. We demonstrate that displacing the PHSP scoring plane 1 mm downstream of the physical aperture exit is both necessary and sufficient to resolve all artifacts across the full clinical range of Mobetron UHDR aperture configurations, yielding a validated applicator-specific PHSP library spanning twelve clinical configurations (2.5–10 cm diameter, A6 and A10 series) suitable for immediate clinical deployment. In addition to the clinical implementation contribution, we characterize the physical basis of the artifact in Geant4's condensed history transport step behavior — a mechanism that, to our knowledge, has not been previously described in the electron PHSP literature and that warrants investigation in any Geant4-based framework scoring phase space distributions at material termination surfaces.

## 2. Methods

### 2.1 Mobetron UHDR System and Applicator Configurations

All simulations were based on the Mobetron UHDR (IntraOp Medical Corp.), a dedicated mobile electron linear accelerator designed for UHDR radiotherapy. Field shaping is accomplished through interchangeable applicator cones and circular aperture inserts that define the treatment field at the patient surface.

Applicator configurations follow a naming convention encoding both cone geometry and field size. The prefix A6 or A10 denotes the applicator cone series — a 19 cm short cone and 28 cm long cone with maximum field sizes of 6 and 10 cm diameter, respectively (Figure 1). The suffix I followed by a numerical value identifies the circular aperture insert diameter in centimeters. This study evaluated the full clinical range: seven A6 configurations spanning 2.5–6.0 cm diameter and five A10 configurations spanning 6.0–10.0 cm diameter, for a total of twelve aperture configurations. This naming convention also establishes the direct aperture-to-PHSP correspondence central to the clinical workflow described here.

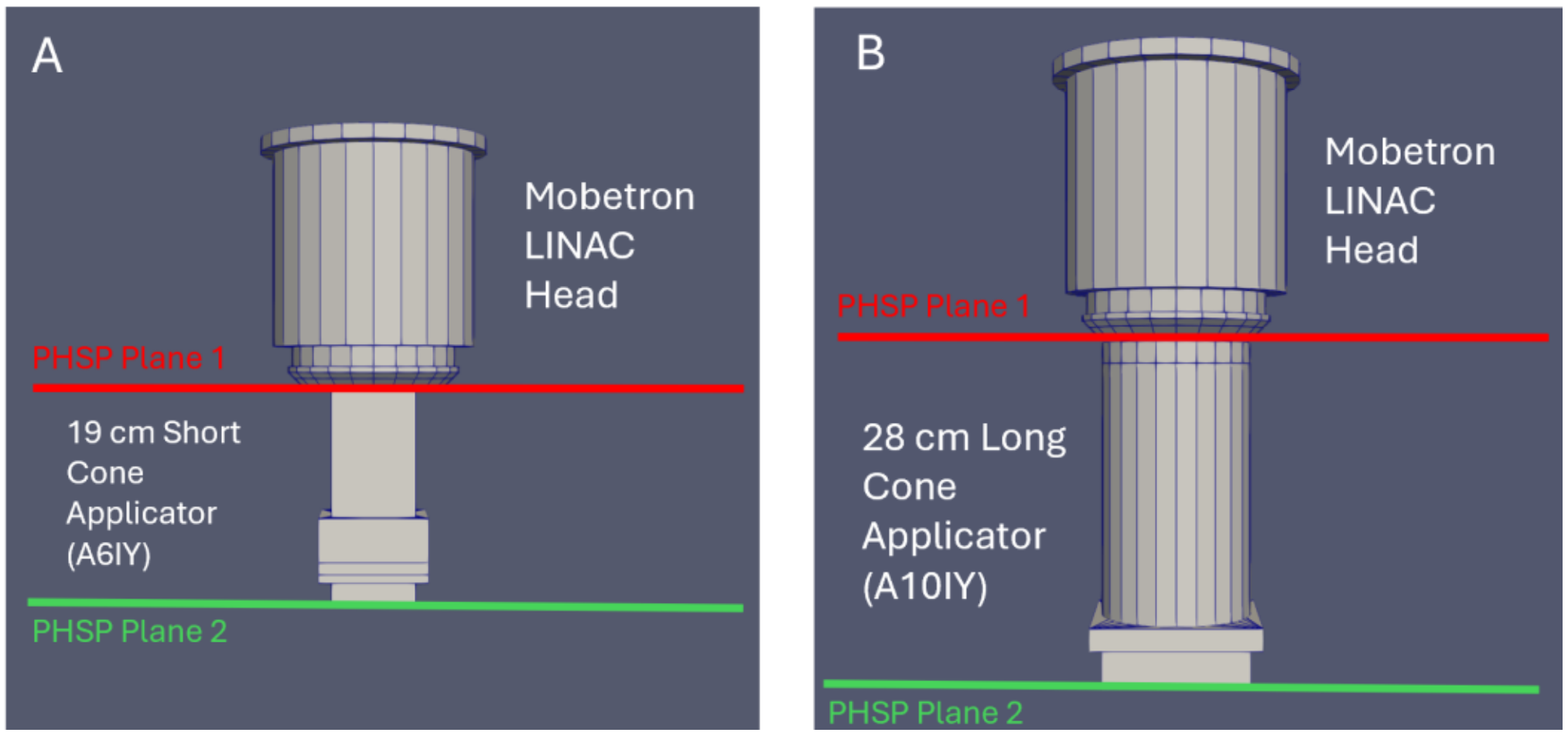

**Figure 1A-B:** Mobetron UHDR LINAC head visualizations for A) the 19 cm short cone applicator (A6IY) and B) the 28 cm long cone applicator (A10IY). Red lines indicate the Linac Exit PHSP plane and green lines indicate the Applicator Exit PHSP plane.

**2.2 Monte Carlo Simulation Framework**

Simulations were performed using GAMOS version 6.2.0, a Geant4-based Monte Carlo platform for medical physics and radiation transport applications, built on Geant4 version 10.6.p02 as bundled with the GAMOS 6.2.0 distribution[5]. The full Mobetron treatment head geometry was implemented via the GAMOS text geometry interface, explicitly modeling all beam-shaping components including the primary collimator, scattering foils, monitor chamber, secondary collimator, and applicator cone assembly for all twelve configurations following a previously published MC model[3]. The aperture insert material in the simulation geometry is ACETAL (polyoxymethylene, density 1.41 g/cm$^3$, composed of $C_6H_{14}O_2$), a low-Z acetal resin with $Z_{eff} \approx 5.7$.

Beam source parameters were derived from independent commissioning measurements: a Gaussian energy distribution (mean 9.58 MeV, sigma 0.173 MeV) and a Gaussian spatial distribution (sigma 1 mm) along the central axis. Particle transport was performed using the GmEMPhysics constructor, corresponding to Geant4's G4EmStandardPhysics (Option 0)[4,5,6]. While GmEMPhysics/opt0 is not the standard physics list choice for medical physics applications — which typically use opt3 or opt4 (equivalent to GAMOS's GmRadiotherapyPhysics with fUseSafetyPlus and Skin = 3) — it was selected here because its default fUseSafety step-limitation algorithm cleanly exposes the boundary sampling behavior under investigation. The performance of opt4-equivalent physics (fUseSafetyPlus with Skin = 3) at the same scoring plane position is characterized as a comparison case in Section 3.1 and shows partial mitigation of the artifact. This constructor employs the G4UrbanMscModel with the fUseSafety step-limitation algorithm, in which maximum step length is constrained by:

$$L = \max\{fR \times R_i, fS \times safety\} \qquad 1$$

where fR = 0.04 is a range factor, $R_i$ is the particle's remaining range, fS = 0.6 is a safety factor, and safety is the isotropic distance to the nearest geometry boundary[8]. The skin parameter is 0, meaning single-scattering mode is never activated near boundaries. The behavior of this step algorithm at exact material boundaries — where safety = 0 — is central to understanding the artifacts investigated in this work and is discussed in detail in Section 4.1.

**2.3 Phase Space File Generation and Boundary Offset Investigation**

Applicator-specific PHSP files were generated using the GAMOS RTPhaseSpaceUA user action, which records particle state at a defined axial Z position. All forward-directed particles were recorded without energy or angular filtering, capturing the complete population including primary electrons, secondary electrons, and bremsstrahlung photons. Particles were not terminated at the scoring plane (KillAfterLastZStop FALSE), ensuring downstream transport was unaffected by the recording. Files were stored in IAEA standard format with a maximum of $5 \times 10^8$ particles per file. For this work, $10^8$ particles were scored per PHSP file — a value determined by preliminary convergence testing to yield statistical uncertainty below 1-2% while maintaining practical simulation times. Downstream water phantom dose calculations used $10^6$ particles per calculation.

The central variable of this investigation was the axial position of the scoring plane relative to the physical aperture exit. This exit surface is located at 397 mm from the electron source for the A6 short cone (19 cm) and at 406 mm for the A10 long cone (28 cm), reflecting the different cone lengths. Because the artifact is expected to be most severe for the smallest apertures (Section 4.3), the sensitivity analysis was performed on the A6 configuration across three positions:

- 397 mm — coincident with the aperture material-air interface (baseline failure configuration)
- 397.1 mm — 0.1 mm downstream (sensitivity analysis)
- 398 mm — 1 mm downstream (corrected configuration)

The corresponding 1 mm offset was applied to the A10 long cone at 407 mm for all production PHSP files used in the global validation (Section 3.3). Baseline (no-offset) PHSP generation was not repeated for the A10 configurations because the artifact magnitude for large apertures is expected to be negligible on physical grounds and the corrected A10 PHSP files showed excellent agreement with measured data in the validation analysis.

Reference PHSP files were generated at the linac exit window position (196 mm from source) under identical simulation conditions. At this distance, the scoring plane intersects the open air bore along the central axis of the applicator assembly which is surrounded by structural components at larger radii but not coincident with any material surface on axis. Safety values are nonzero for all on-axis particles at this position, and the fUseSafety algorithm functions normally. This represents standard PHSP scoring practice and serves as the ground-truth benchmark for all comparisons[9].

### 2.4 Dose Calculation and Experimental Validation

Dose calculation was performed in a 48 x 48 x 41cm water phantom with a 2 cm air gap between the applicator exit and the phantom entrance, utilizing a 1 mm voxel resolution for dose scoring. Experimental validation was conducted using an IBA Blue Phantom water scanning system equipped with a Flash Diamond field detector and IC-15 3573 reference detector for depth-dose measurements. Spatial accuracy between the simulated and measured data was quantified using distance-to-agreement (DTA) analysis, with a clinical acceptance criterion of <3 mm[7]. Simulations were performed on an Intel Core i9-13980HX (24 cores) workstation with 64 GB RAM. Typical PHSP generation required 15–20 minutes per file at $10^8$ particles, and water phantom dose calculations completed in a few minutes at $10^6$ particles per calculation.

## 3. Results

### 3.1 PHSP Plane Boundary Artifact Identification, Sensitivity, and Mitigation

Initial attempts to generate PHSP files with the scoring plane coincident with the physical aperture exit at 397 mm produced significant dosimetric discrepancies across all configurations tested. The A6I3 configuration is shown as a representative case in Figure 2. With the scoring plane at the exact collimator

- air interface, the PDD exhibited a premature descent on the falloff slope and a proximal shift in R50 from the measured value of 3.42 cm to 3.20 cm corresponding to a 2.2 mm error (Figure 2A). DTA analysis across the full PDD curve yielded values of 4–6 mm for the smallest aperture configurations, well outside the 3 mm clinical acceptance criterion. Displacing the scoring plane 1 mm distal to the boundary at 398 mm fully restored the PDD, recovering R50 to 3.40 cm and matching measured data across the buildup, falloff, and distal regions (Figure 2A).

To characterize the minimum offset distance required for full transport stabilization, the scoring plane was evaluated at intermediate positions of 397.1 mm (0.1 mm offset) and 398 mm (1 mm offset), with the linac-exit reference scored at 196 mm for comparison (Figure 2B). The 0.1 mm offset produced partial recovery: R50 shifted from 3.20 cm to 3.30 cm, and the steepness of the falloff slope improved, indicating that the primary electron energy spectrum began to stabilize. However, the distal falloff region remained systematically underrepresented relative to both the linac-exit reference and measured data. The 1 mm offset (398 mm) resolved both the falloff slope and the distal region simultaneously with R50 = 3.40 cm vs measured 3.42 cm, achieving dosimetric equivalence to the linac-exit reference. The dissociation between R50 recovery (0.1 mm sufficient) and full distal falloff recovery (requires 1 mm) is characterized as a two-stage recovery hierarchy, with the underlying mechanism discussed in Section 4.1.

To investigate whether alternative MSC configurations could mitigate the artifact without a geometric offset, simulations were repeated at the exact interface (397 mm) using the GmRadiotherapyPhysics physics list, which implements the fUseSafetyPlus step-limitation algorithm with Skin = 3 — activating explicit single scattering within three mean free paths of any geometric boundary (Figure 2C). This configuration produced partial recovery equivalent to the 0.1 mm geometric offset: R50 was restored from 3.20 cm to 3.30 cm, and the buildup region was substantially improved, but the distal falloff deficit remained unresolved. The Skin = 3 result was indistinguishable from the 0.1 mm GmEMPhysics offset across the full depth range. Full recovery of both the R50 and distal falloff required the 1 mm geometric offset regardless of MSC configuration. The physical mechanism underlying these observations is discussed in detail in Section 4.1.

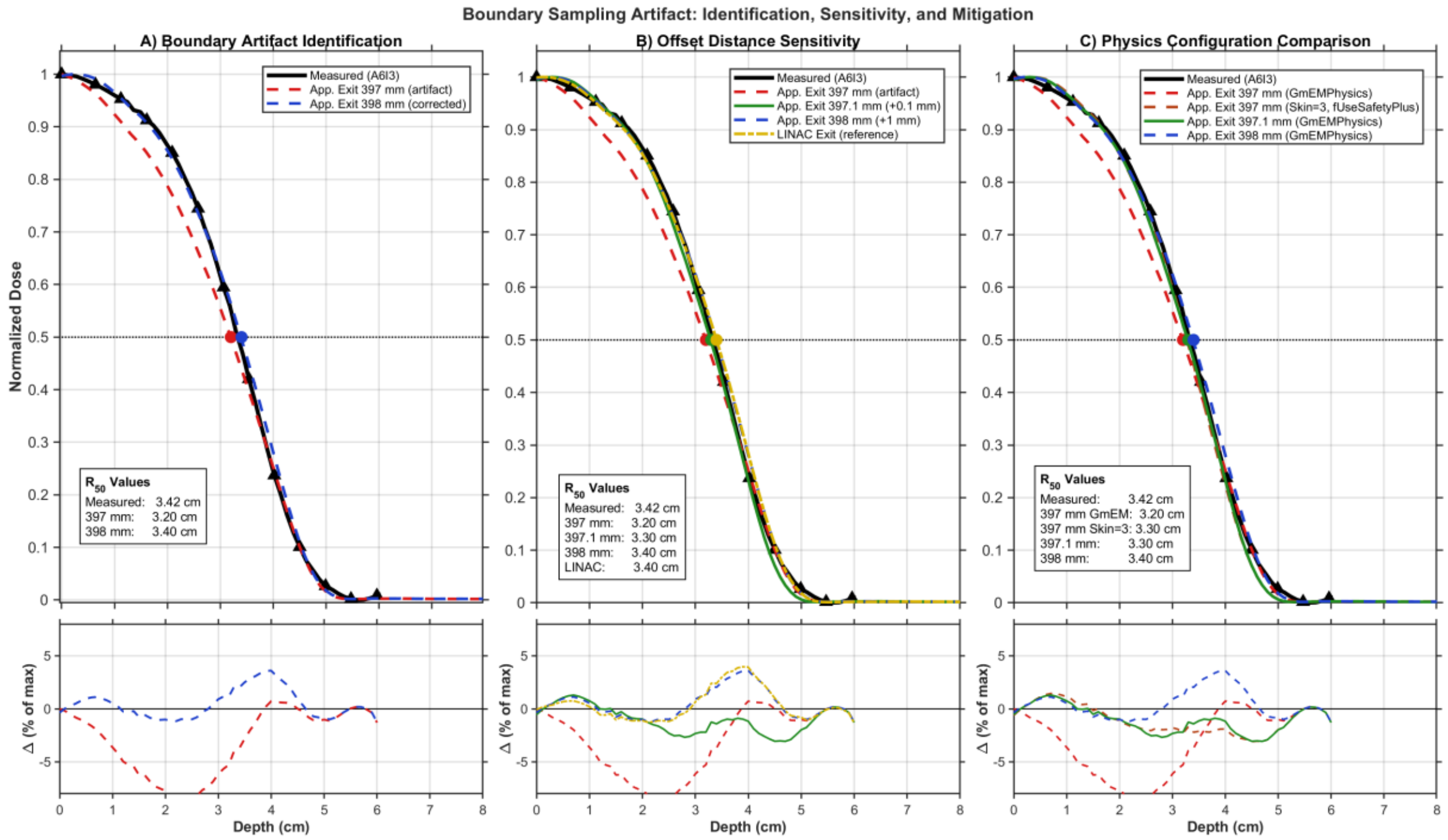

**Figure 2:** Boundary sampling artifact behavior for the representative A6I3 configuration. Upper panels show percent depth dose (PDD) curves; lower panels show local dose residuals (simulation minus measured, expressed as percent of maximum dose) as a function of depth. A) Identification of the artifact: the PHSP scored at the exact material-air interface (397 mm, red) produces a 2.2 mm proximal R50 shift and premature PDD falloff relative to measured data (black), while a 1 mm downstream offset (398 mm, blue) restores agreement. B) Offset distance sensitivity: 0.1 mm offset (green) partially recovers R50 but retains a sustained negative residual of 2–3% throughout the distal falloff region (approximately 3–6 cm depth), while the 1 mm offset (blue) and linac-exit reference (gold) both achieve full recovery with residuals distributed around zero. C) Physics configuration comparison: scoring at 397 mm with GmRadiotherapyPhysics (Skin = 3, fUseSafetyPlus, dark orange) produces a residual pattern indistinguishable from the 0.1 mm GmEMPhysics offset (green including the same 2–3% distal falloff deficit; only the 1 mm geometric offset (blue) eliminates this systematic bias.

Residual analysis (Figure 2, lower panels) quantifies these three regimes directly. The 397 mm baseline configuration produces a systematic negative excursion reaching −8% of maximum dose concentrated in the buildup and falloff region (0–3 cm depth), corresponding to the boundary-truncated secondary electron population. The 0.1 mm offset and Skin = 3 configurations recover the buildup and falloff region to within ±2% but exhibit a sustained negative residual of 2–3% across the distal falloff and near-Rp region (approximately 3–6 cm depth), reflecting the residual energy-loss artifact that partial mitigation does not fully address. In contrast, the 1 mm offset and linac-exit reference show residuals distributed around zero without systematic depth-dependent bias, with residual excursions consistent with the 1 mm dose scoring voxel resolution and setup registration uncertainty. The distinction between the systematic distal deficit of the partially mitigated configurations and the zero-centered residual pattern of the 1 mm offset is the key qualitative signature that establishes the 1 mm offset as the fully corrective configuration rather than a marginally better version of the partial mitigation.

### 3.2 Angular Phase Space Analysis

To further characterize the phase space differences underlying the observed dosimetric recovery, angular distribution analysis was performed on the PHSP files generated for the representative A6I3 configuration at the 0mm and 1mm scoring plane positions. Energy spectra comparison between the two offsets revealed negligible differences across the full energy range, consistent with the artifact affecting a small secondary particle sub-population rather than the bulk energy distribution. The angular distribution, however, revealed a clear and physically interpretable signature of the boundary sampling artifact.

To quantify the difference, the angular frequency difference between the 1mm and 0mm offset PHSP files was computed by binning all scored particles by angle with respect to the beam axis and subtracting the normalized 0mm distribution from the 1mm distribution (Figure 3). At angles below approximately 5°, the 0mm offset PHSP contains an excess of particles relative to the 1mm offset (negative ΔFrequency), manifesting as a numerical pileup of near-forward particles artificially concentrated on-axis. At angles between approximately 10° and 90°, the relationship inverts: the 1mm offset PHSP contains a consistent excess of particles (positive ΔFrequency), corresponding to the large-angle secondary electron population absent from the 0mm PHSP due to boundary truncation. The mechanism responsible for this angular redistribution is discussed in Section 4.1.

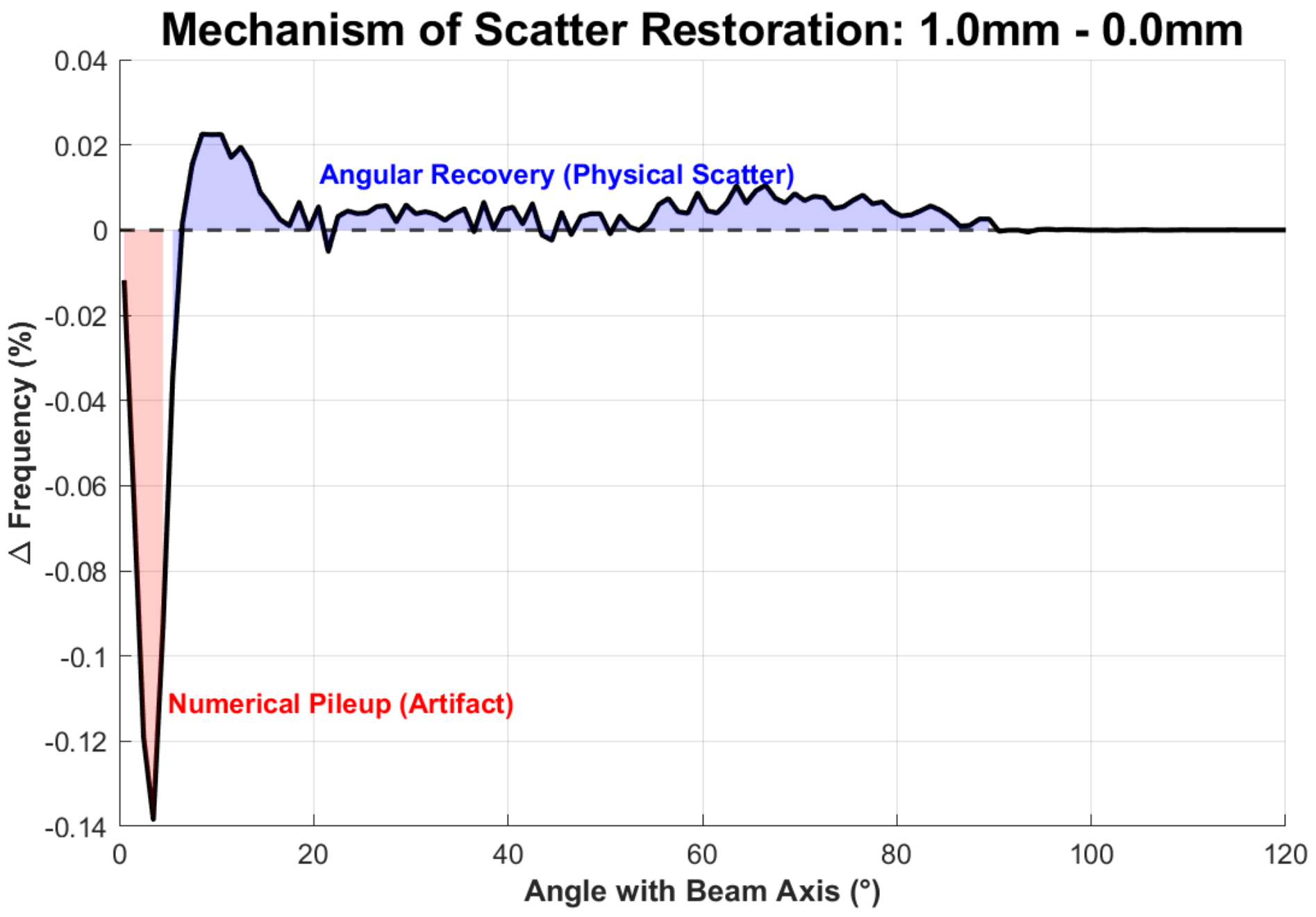

**Figure 3:** Angular frequency difference ($\Delta f(\theta) = f_{1mm}(\theta) - f_{0mm}(\theta)$) between the 1mm and 0mm offset applicator-exit PHSP files for the A6I3 configuration. The red shaded region (0–~8°) identifies the numerical pileup artifact — a near-forward particle excess in the 0mm PHSP. The blue shaded region (~10–90°) represents the recovered large-angle secondary electron population present in the 1mm offset PHSP but absent from the 0mm PHSP due to boundary truncation.

**3.4 Distance-to-Agreement Analysis**

To confirm that the 1 mm offset generalizes, applicator-exit PHSP files with the correction were generated for all twelve aperture configurations (seven A6, five A10) and compared with both linac-exit PHSP simulations and measured water-scan data. PDDs and lateral profiles at $d_{max}$ for every configuration are shown in Supplementary Figures S1, S2 (A6) and S3 (A10). Applicator-exit results were dosimetrically equivalent to linac-exit results and measured data across all configurations, and this agreement is quantified by DTA analysis in Figure 4.

Quantitative spatial accuracy was assessed via DTA analysis across all twelve configurations and both PHSP generation methods (Figure 4). For the A6 series, applicator-exit PHSP files achieved mean PDD DTA values of 1.45 ± 0.63 mm (range: 0.60–2.29 mm) and mean lateral profile DTA values of 1.64 ± 1.34 mm (range: 0.50–3.66 mm). Linac-exit PHSP files performed comparably, with mean PDD DTA values of 1.55 ± 0.51 mm (range: 0.94–2.44 mm) and mean profile DTA values of 1.47 ± 1.36 mm (range: 0.50–3.66 mm). For the A10 series, applicator-exit PHSP files yielded mean PDD DTA values of 2.00 ± 0.61 mm (range: 1.21–2.64 mm) and mean profile DTA values of 1.75 ± 0.97 mm (range: 1.01–3.39 mm), while linac-exit files yielded mean PDD DTA values of 2.42 ± 0.86 mm (range: 1.38–3.21 mm) and mean profile DTA values of 2.01 ± 0.79 mm (range: 1.43–3.02 mm). All mean DTA values were well within the 3 mm clinical acceptance criterion.

Notably, applicator-exit PHSP files achieved equal or lower mean DTA values than linac-exit files for both the A6 and A10 PDD comparisons, confirming that eliminating redundant applicator transport does not degrade and in some configurations marginally improves spatial accuracy relative to the traditional approach. Figures 4C and 4D show per-aperture mean DTA values remaining below the 3 mm clinical acceptance criterion across all configurations for both PHSP generation methods, with applicator-exit values equal to or lower than linac-exit values in all A6 and A10 PDD comparisons. The magnitude of residual DTA values across configurations is consistent with the 1 mm dose grid voxel resolution rather than with systematic model error.

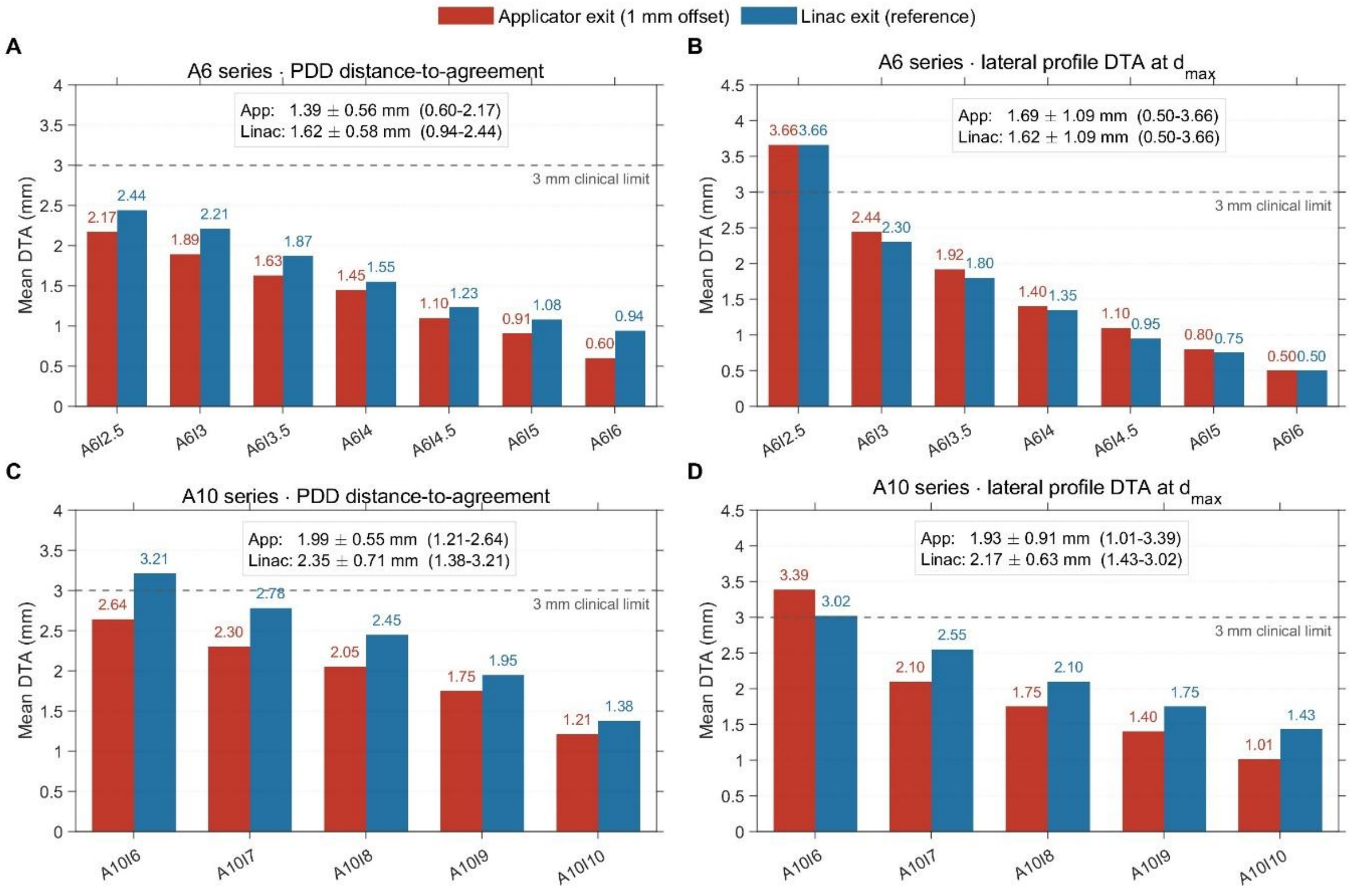

**Figure 4A–D:** Distance-to-agreement analysis across all twelve aperture configurations, comparing applicator-exit PHSP files (1 mm offset, red) against linac-exit reference PHSP files (blue). **(A)** Mean PDD DTA for the seven A6 series apertures (2.5–6.0 cm). **(B)** Mean lateral profile DTA at $d_{max}$ for the A6 series. **(C)** Mean PDD DTA for the five A10 series apertures (6.0–10.0 cm). **(D)** Mean lateral profile DTA at $d_{max}$ for the A10 series. Per-aperture values are labeled above each bar; inset boxes report the series-level mean ± SD with min–max range. The dashed gray line indicates the 3 mm clinical acceptance criterion. All twelve configurations satisfy the criterion for both PHSP generation methods, with applicator-exit DTA values equal to or lower than linac-exit values for the PDD comparisons in both series.

## 4. Discussion

### 4.1 Physical Mechanism of Boundary Sampling Artifacts

Geant4 simulates MeV-range electron transport using condensed history methods in which the G4UrbanMscModel approximates thousands of individual elastic scattering events into discrete transport steps governed by Lewis multiple scattering theory. This formalism assumes an infinite homogeneous medium over each step length — an assumption violated at material boundaries — and the consequences are well documented[10]: condensed history algorithms underestimate lateral displacement near interfaces and permit electrons to reach geometric boundaries in inappropriately large steps, producing dose errors of up to 39% in cavity geometries[10,11]. The present work demonstrates that these same boundary-crossing artifacts corrupt PHSP files when the scoring plane is coincident with the interface.

The mechanism operates through three coupled failure modes. First, when a particle sits exactly on a material boundary, Geant4's safety parameter — the isotropic distance to the nearest geometry surface computed by G4Navigator — evaluates to zero. For the fUseSafety step limitation algorithm used by GmEMPhysics, the step limit is L = max{fR × R_initial, fS × safety}. With safety = 0, the proximity constraint vanishes, and the algorithm defaults to fR × R_initial alone, which at 9 MeV in ACETAL may permit a first step in air of several millimeters without proper path-length correction. This degenerate condition is acknowledged in the Geant4 source code (G4CoupledTransportation): no safety update is performed when a particle is at a boundary[6,8].

Second, the final transport step within the aperture material is always truncated at the interface. Multiple scattering models such as the G4UrbanMscModel compute angular deflection and lateral displacement as statistical accumulations over a sampled step length in a homogeneous medium — quantities whose variance scales with the path over which they are integrated. In the Lewis formulation employed by G4UrbanMscModel, the mean-square lateral displacement grows approximately as $s^2$ for a given angular distribution and cannot be self-consistently recovered when the step is foreshortened at a boundary. The nontrivial consequence is not that a shorter step produces less lateral spread — which is trivially true — but that truncated-step transport underestimates the lateral spread that properly resolved transport (with physics resampled over multiple sub-steps, as employed by EGSnrc's exact boundary crossing algorithm[12,13]) would have produced at the same geometric exit point. Unlike EGSnrc, which switches to explicit single scattering within a defined distance of any boundary, the default fUseSafety algorithm in GmEMPhysics does not activate single scattering near interfaces (Skin = 0), so the boundary-truncated step is simply executed with its foreshortened physics accumulation. Low-energy secondary electrons generated at the aperture wall undergo additional scattering before exiting the material; when the step is truncated at the exact boundary, these secondaries are clipped from the transport history before their trajectories stabilize in air. They are therefore absent from or incorrectly represented in the scored PHSP, producing a deficit of low-energy, large-angle particles responsible for buildup dose deposition. The proximal R50 shift and premature PDD falloff are the direct dosimetric signatures of this secondary electron deficit. This angular redistribution is directly visualized in Figure 3, which shows the near-forward particle excess in the 0 mm PHSP (0–8°) and the corresponding deficit of large-angle secondaries (10–90°) that are recovered with the 1 mm offset.

Third, the truncated final step in ACETAL introduces a systematic energy-loss artifact in the scored particle population. Condensed history transport samples energy loss and path-length corrections for the intended step length assuming a homogeneous medium; when the step is terminated at the material-air boundary, the correspondence between the sampled physics and the actually traversed path is disrupted, and the last increment of transport carries material-appropriate physics parameters into the air region. The scored particles emerge with a mean kinetic energy that is systematically lower than would have been recorded had they completed their transition to air-appropriate transport statistics before scoring. The dosimetric consequence is a proximal shift of the entire distal falloff region, including the near-Rp region, which manifests in Figure 2 as a sustained 2–3% dose deficit at depths of 3–6 cm in the uncorrected and partially-corrected configurations. Notably, this deficit is not a suppression of true bremsstrahlung contamination — which for a 9 MeV beam originates predominantly upstream in the scattering foils and beam-line components and appears as a slowly-diminishing horizontal tail beyond Rp of approximately 1–2% of Dmax — but is instead a shift-induced deficit in the falloff region driven by the reduced effective beam energy at scoring.

This energy-loss failure mode is distinct from the angular redistribution above and operates on a longer length scale, as confirmed by the comparison in Figure 2C. Activating fUseSafetyPlus with Skin = 3 — which forces explicit single scattering within three mean free paths of any boundary — produced recovery equivalent to the 0.1 mm geometric offset, matching both the R50 shift and the persistent 2–3% distal falloff deficit. Both mitigations address the angular pileup at the boundary but neither provides sufficient transport distance in air for the scored particle population to fully re-equilibrate to air-appropriate transport statistics before being recorded. Only the 1 mm geometric offset resolves both the angular and energy-loss components simultaneously.

An important geometric distinction explains why this artifact is specific to the applicator-exit scoring position and does not affect conventional linac-exit phase space scoring. The affected particle population is not on-axis primary electrons — which traverse the open aperture in air — but rather the wall-scattered secondary electrons emerging from the ACETAL aperture surface. The critical difference between the two scoring positions is the availability of downstream transport geometry for this wall-scattered population to equilibrate before being recorded. At 196 mm, wall-scattered electrons from upstream beam-line components still have several centimeters of downstream applicator cone and aperture insert to traverse before reaching the phantom, providing full path length for their trajectories to stabilize under the appropriate transport statistics. At 397 mm, wall-scattered electrons emerging from the ACETAL surface are captured at the instant they cross the material-air interface — safety = 0 for these particles at the moment of scoring, and there is no downstream geometry remaining to equilibrate their transport before recording. The artifact therefore arises specifically when a scoring plane coincides with a collimator or aperture exit surface, capturing wall-scattered secondary electrons before they have completed their transition to air-appropriate transport. This mechanism is consistent with the aperture size dependence characterized in Section 4.3, where smaller apertures produce a larger wall-scattered fluence fraction and correspondingly larger artifact magnitude. Notably, the ACETAL aperture is a low-Z material rather than a high-Z metal, so the artifact appears despite the reduced secondary particle yield relative to metallic collimators — confirming the driver is the boundary step geometry rather than the secondary yield magnitude. The concern therefore extends to any scoring plane coincident with a collimator or aperture exit face regardless of atomic number, wherever wall-scattered secondaries are captured without downstream equilibration.

### 4.2 The 1 mm Recovery Distance

The two-stage recovery hierarchy identified in Section 3.1 reflects the distinct length scales required for the two independent physical processes described in Section 4.1. Primary electron angular equilibration after the final boundary-truncated step is recovered by any mechanism that provides even sub-millimeter transport distance in air or activates single scattering near the boundary, as evidenced by the equivalence of the 0.1 mm geometric offset and the Skin = 3 physics configuration. Full recovery of the distal falloff has a larger characteristic length scale. The scored particle population must complete the transition from material-appropriate to air-appropriate condensed-history transport before being recorded, which requires sufficient in-air path length for the boundary-crossing step and the first in-air step to resample under air-appropriate physics. At 9 MeV in air, typical condensed-history step lengths are of order 0.5–1 mm[15], and low-energy secondary electrons (10–100 keV) generated near the boundary have ranges of roughly 0.1–1 mm in air[14]; together these establish 1 mm as the threshold at which the sampled transport statistics correspond to the medium at scoring. That improved MSC step handling alone cannot recover the distal falloff confirms this is fundamentally a transport-distance problem rather than a step-calculation problem — boundary-proximal step algorithms cannot substitute for the physical path length required for the sampled physics to correspond to the medium. The 1 mm offset introduces negligible perturbation to the primary beam: energy loss across 1 mm of air at 9 MeV is under 0.01% and angular spread changes are below 0.05°.

### 4.3 Aperture Size Dependence

The severity of the boundary artifact scales inversely with aperture diameter. For the 2.5 cm A6I2.5 configuration, DTA errors reach 4–6 mm; for larger apertures the errors diminish. This reflects the geometric relationship between the open-field and wall-scattered particle populations. For small apertures, the perimeter-to-area ratio is high and a large fraction of the beam fluence consists of particles that have interacted with the collimator wall, which is the population most severely affected by all three failure modes. For large apertures, most particles traverse the open field without wall interaction, and boundary artifacts affect only a small fraction of total fluence. The practical implication is that boundary sampling artifacts are most dangerous precisely in the configurations where treatment fields are smallest and spatial precision is most critical.

**4.4 Clinical Implementation and Generalizability**

The applicator-specific PHSP library developed and validated in this work is being deployed in the GMV Radiance treatment planning system as the foundational source model for Mobetron UHDR FLASH-RT commissioning. Characterization of the boundary sampling artifact was a prerequisite for this deployment: the uncorrected R50 shifts of up to 2.2 mm and DTA errors of 4–6 mm exceeds the AAPM TG-105 clinical acceptance criterion[7] and would preclude clinical use of the library. The 1 mm offset correction (mean PDD DTA 1.45 ± 0.63 mm for A6, 2.00 ± 0.61 mm for A10) resolves the artifact and supports downstream TPS commissioning.

Beyond the immediate Mobetron application, the 30–50% reduction in dose calculation time achieved by applicator-specific PHSP files addresses a specific barrier to clinical FLASH-RT deployment, where UHDR delivery timelines demand faster planning cycles than traditional linac-exit workflows can support. The boundary offset requirement itself extends to any Geant4-based simulation in which a PHSP scoring plane coincides with an on-axis material termination surface: the condensed history behavior described in Section 4.1 arises from Geant4's core tracking architecture (G4CoupledTransportation, G4UrbanMscModel), which is shared by TOPAS[16], GATE, and direct Geant4 applications. Investigation in these frameworks is warranted for any clinical MC commissioning workflow employing similar scoring configurations.

**4.4.1 Practical Implementation Guidance**

PHSP scoring planes should be positioned in air at least 1 mm downstream of any material surface when simulation geometry permits, as this resolves both the angular equilibration and energy-loss components of the artifact. Before committing to a production PHSP library for a new geometry, a sensitivity analysis comparing R50 and distal falloff shape across at least three offset positions (interface, 0.1 mm, 1 mm) is recommended, with the smallest aperture configuration tested first as the most sensitive indicator of residual artifact. Where a 1 mm offset is not geometrically feasible, activating fUseSafetyPlus with Skin = 3 (available via GmRadiotherapyPhysics in GAMOS or G4EmStandardPhysics_option4 in native Geant4[5,6]) provides partial mitigation of the boundary angular artifact but leaves the distal falloff deficit unresolved; users should weigh the residual distal dose error against the geometric constraint when full offset is infeasible.

## 5. Conclusion

This work characterizes and validates the mitigation of a boundary sampling artifact in Geant4-based Monte Carlo phase space generation, yielding an applicator-specific PHSP library ready for clinical FLASH-RT deployment. Scoring at the exact applicator–air interface produced systematic R50 shifts of up to 2.2 mm and DTA errors of 4–6 mm for small-aperture configurations, exceeding the AAPM TG-105 clinical acceptance criterion[7] and rendering the uncorrected library unsuitable for deployment. A 1 mm downstream offset of the PHSP scoring plane fully resolves the artifact across all twelve Mobetron aperture configurations tested (mean PDD DTA 1.45 ± 0.63 mm for A6, 2.00 ± 0.61 mm for A10), with negligible perturbation to primary beam characteristics. Partial mitigation via improved MSC step handling (fUseSafetyPlus with Skin = 3) recovers primary electron angular equilibration at the boundary but leaves a persistent 2–3% deficit in the distal falloff region from an energy-loss artifact that requires the full 1 mm offset to resolve, establishing the geometric offset as uniquely sufficient for full clinical-grade recovery. The corrected PHSP library is being deployed in the GMV Radiance TPS for Mobetron FLASH-RT commissioning. Because the underlying mechanism arises from Geant4's core condensed history transport behavior at material boundaries, the boundary offset requirement warrants investigation in any Geant4-based framework — including TOPAS and GATE — being commissioned for MC-based clinical dose calculation.

## References

1. Favaudon V, Caplier L, Monceau V, et al. Ultrahigh dose-rate FLASH irradiation increases the differential response between normal and tumor tissue in mice. Sci Transl Med. 2014;6(245):245ra93. doi:10.1126/scitranslmed.3008973
2. Wilson JD, Hammond EM, Higgins GS and Petersson K (2020) Ultra-High Dose Rate (FLASH) Radiotherapy: Silver Bullet or Fool's Gold?. Front. Oncol. 9:1563. doi: 10.3389/fonc.2019.01563

3. Dai T, Sloop AM, Rahman MR, et al. First Monte Carlo beam model for ultra-high dose rate radiotherapy with a compact electron LINAC. Med Phys. 2024;51:5109–5118. https://doi.org/10.1002/mp.17031
4. S. Agostinelli et al., Geant4—a simulation toolkit, Nuclear Instruments and Methods in Physics Research Section A: Accelerators, Spectrometers, Detectors and Associated Equipment, Volume 506, Issue 3, 2003, Pages 250-303, ISSN 0168-9002, https://doi.org/10.1016/S0168-9002(03)01368-8.
5. P. Arce, J.I. Lagares, L. Harkness, D. Pérez-Astudillo, M. Cañadas, P. Rato, M. de Prado, Y. Abreu, G. de Lorenzo, M. Kolstein, A. Díaz, Nucl. Instrum. Methods Phys. Res. A 735 (2014) 304–313. https://doi.org/10.1016/j.nima.2013.09.036
6. Allison J, Amako K, Apostolakis J, et al. Recent developments in Geant4. Nucl Instrum Methods Phys Res A. 2016;835:186-225. doi:10.1016/j.nima.2016.06.125
7. Chetty, I.J., Curran, B., Cygler, J.E., DeMarco, J.J., Ezzell, G., Faddegon, B.A., Kawrakow, I., Keall, P.J., Liu, H., Ma, C.-.-M.C., Rogers, D.W.O., Seuntjens, J., Sheikh-Bagheri, D. and Siebers, J.V. (2007), Report of the AAPM Task Group No. 105: Issues associated with clinical implementation of Monte Carlo-based photon and electron external beam treatment planning. Med. Phys., 34: 4818-4853. https://doi.org/10.1118/1.2795842.
8. Ivanchenko V, Apostolakis J, Bagulya A, et al. Recent Improvements in Geant4 Electromagnetic Physics Models and Interfaces. Progress in Nuclear Science and Technology. 2011;2:898-903. doi:10.15669/pnst.2.898
9. Capote R, Jeraj R, Ma CM, et al. Phase-space database for external beam radiotherapy. IAEA-TECDOC-1540. Vienna: International Atomic Energy Agency; 2006. https://www-pub.iaea.org/MTCD/publications/PDF/te_1540_web.pdf
10. Poon, E. and Verhaegen, F. (2005), Accuracy of the photon and electron physics in GEANT4 for radiotherapy applications. Med. Phys., 32: 1696-1711. https://doi.org/10.1118/1.1895796
11. Bielajew, Alex F., and D.W.O. Rogers. "Presta: The Parameter Reduced Electron-Step Transport Algorithm for Electron Monte Carlo Transport." Nuclear Instruments and Methods in Physics Research Section B: Beam Interactions with Materials and Atoms, vol. 18, no. 1-6, 1986, pp. 165–181, https://doi.org/10.1016/s0168-583x(86)80027-1. Accessed 29 Jan. 2023.
12. Kawrakow, I. (2000), Accurate condensed history Monte Carlo simulation of electron transport. I. EGSnrc, the new EGS4 version. Med. Phys., 27: 485-498. https://doi.org/10.1118/1.598917

13. Kawrakow I, Bielajew AF. On the condensed history technique for electron transport. Nucl Instrum Methods Phys Res B. 1998;142(3):253-280. doi:10.1016/S0168-583X(98)00274-2
14. Berger MJ, Coursey JS, Zucker MA, Chang J. ESTAR, PSTAR, and ASTAR: Computer Programs for Calculating Stopping-Power and Range Tables for Electrons, Protons, and Helium Ions (version 2.0.1). NIST Standard Reference Database 124. National Institute of Standards and Technology; 2017. doi:10.18434/T4NC7P
15. Sempau J, Sánchez-Reyes A, Salvat F, ben Tahar HO, Jiang SB, Fernández-Varea JM. Monte Carlo simulation of electron beams from an accelerator head using PENELOPE. Phys Med Biol. 2001;46(4):1163-1186. doi:10.1088/0031-9155/46/4/318
16. Perl J, Shin J, Schümann J, Faddegon B, Paganetti H. TOPAS: an innovative proton Monte Carlo platform for research and clinical applications. Med Phys. 2012;39(11):6818-6837. doi:10.1118/1.4758060

**Supplementary Material:**

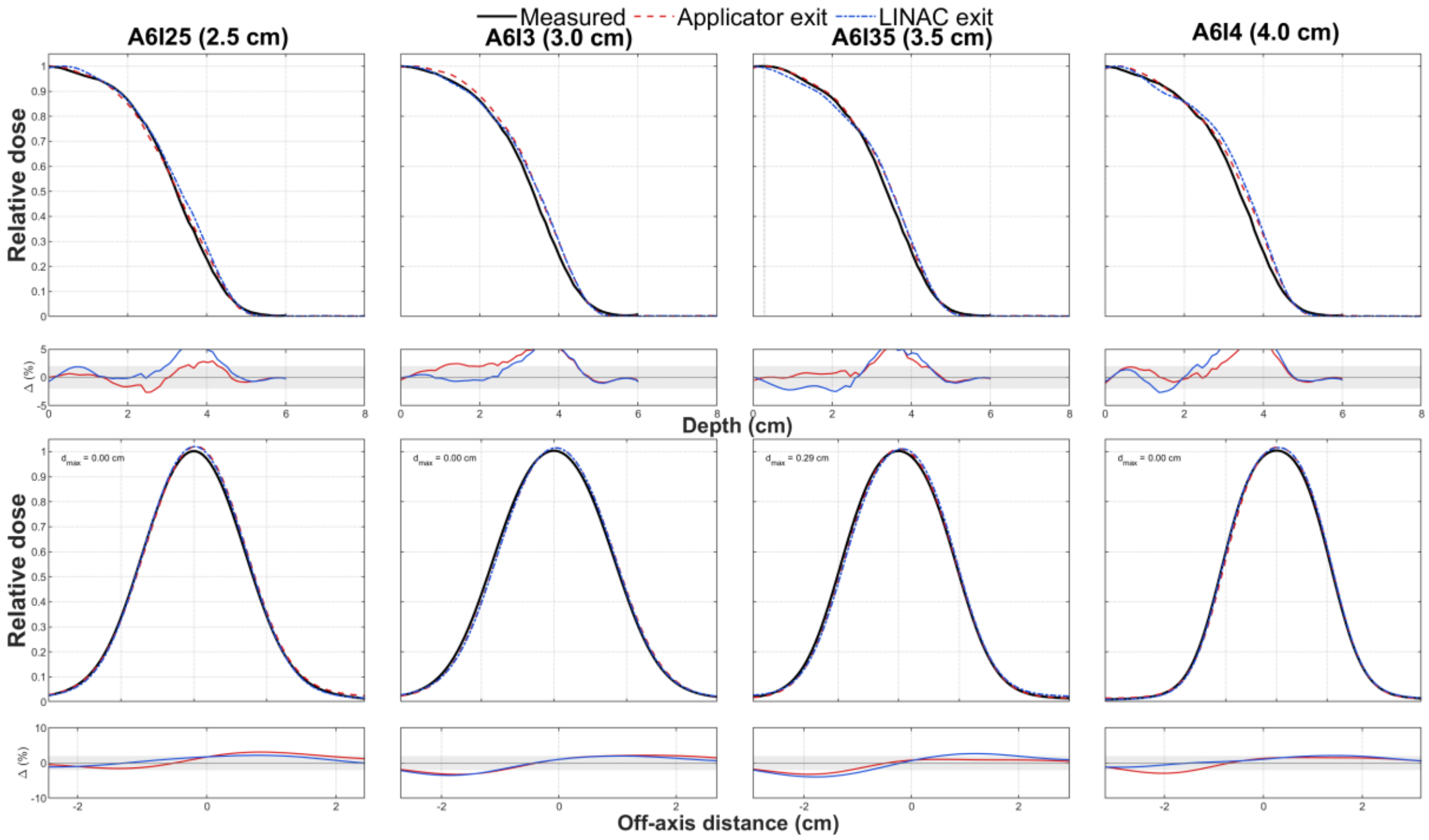

**Figure S1:** A6 series PDDs and lateral profiles for the 2.5–4.0 cm apertures (A6I2.5, A6I3, A6I3.5, A6I4). Columns correspond to individual aperture configurations. Rows show, from top to bottom: PDD; PDD residual; lateral dose profile at $d_{max}$; and profile residual. Measured water-scan data are shown in black (solid), applicator-exit PHSP simulations with the 1 mm scoring plane offset in red (dashed), and linac-exit PHSP simulations in blue (dash-dot). The dotted vertical line marks the measured d~max~. Residuals are MC minus measured, expressed as a percentage of the normalized dose, with the gray band indicating ±2%. Profile residuals are shown only where the measured dose exceeds 10% of the central-axis value; large residuals in the penumbra reflect small lateral shifts on a steep gradient and are quantified by DTA in Figure 4.

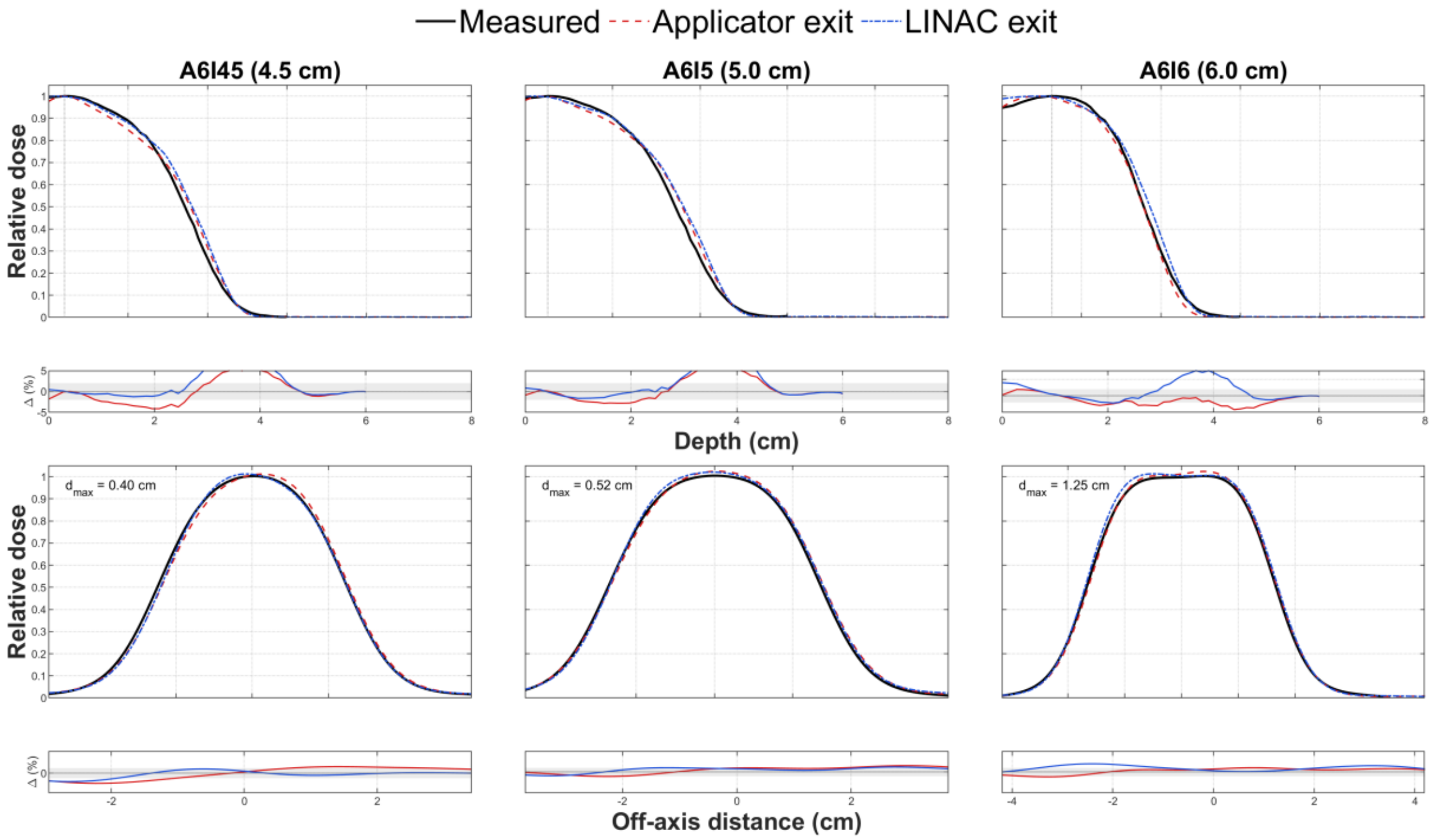

**Figure S2:** A6 series PDDs and lateral profiles for the 4.5–6.0 cm apertures (A6I4.5, A6I5, A6I6). Panel layout, curve definitions, and residual calculations are identical to Figure S1.

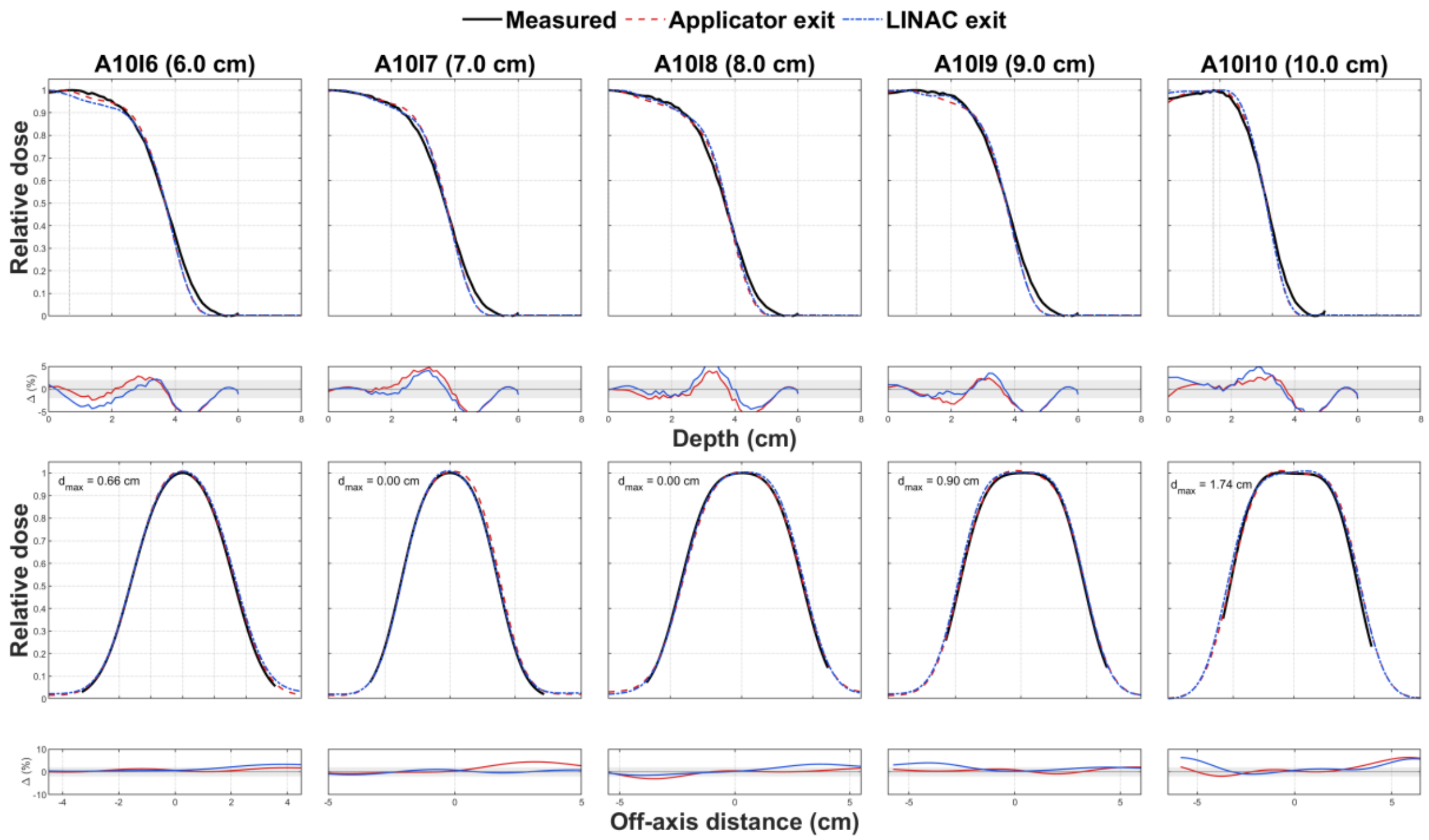

**Figure S3:** A10 series PDDs and lateral profiles for the 6.0–10.0 cm apertures (A10I6, A10I7, A10I8, A10I9, A10I10). Panel layout, curve definitions, and residual calculations are identical to Figure S1.